\documentclass[final,3p,times]{elsarticle}

\usepackage{wrapfig}

\usepackage{graphics}
\usepackage{graphicx}
\usepackage{epsfig}
\usepackage{amssymb}
\usepackage{amsthm}
\usepackage{amsmath}

\usepackage{caption}
\usepackage[labelfont=bf]{caption}
\usepackage[labelsep=space]{caption}
\usepackage[nodots, compress]{numcompress}
\biboptions{sort&compress}

\usepackage{xcolor}
\usepackage{hyperref}
\hypersetup{
colorlinks,
citecolor=[violet],
linkcolor=[red],
urlcolor=[blue]
}

\journal{Next Nanotechnology}

\begin{document}

\begin{frontmatter}

\title{Ultrasonic-assisted liquid phase exfoliation for high-yield monolayer graphene with enhanced crystallinity}

\author[1]{Kaitong Sun\fnref{fn1}}
\author[2]{Si Wu\fnref{fn1}}
\author[3]{Junchao Xia\fnref{fn1}}
\fntext[fn1]{These authors contributed equally.}
\author[1]{Yinghao Zhu}
\author[1]{Guanping Xu}
\author[1]{Hai-Feng Li\corref{cor1}}
\cortext[cor1]{Corresponding author}
\ead{haifengli@um.edu.mo}

\affiliation[1]{organization={Institute of Applied Physics and Materials Engineering, University of Macau},
            addressline={Avenida da Universidade, Taipa},
            city={Macao SAR},
            postcode={999078},
            country={China}}
\affiliation[2]{organization={School of Physical Science and Technology, Ningbo University},
            city={Ningbo},
            postcode={315211},
            country={China}}
\affiliation[3]{organization={School of Materials Science and Engineering, Chang’an University},
            city={Xi’an },
            postcode={710061},
            country={China}}

\begin{abstract}
Graphene stands as a promising material with vast potential across energy storage, electronics, etc. Here, we present a novel mechanical approach utilizing ultrasonic high-energy intercalation exfoliation to extract monolayer graphene from graphite, offering a simple yet efficient alternative to conventional methods. Through a comprehensive series of characterizations involving atomic force microscopy, scanning electron microscopy, Raman spectroscopy, X-ray diffraction, and X-ray photoelectron spectroscopy, the resulting graphene nanosheets demonstrate superior crystallinity compared to those obtained via the conventional method. The high-crystalline freestanding graphene nanosheets derived from this method not only facilitate easier separation but also significantly enhance the physical performance of the original materials. This method showcases the potential for scalable production of layered materials with increased yield and crystallinity, paving the way for their utilization in various applications.
\end{abstract}

\begin{graphicalabstract}
\includegraphics[width=0.88\textwidth]{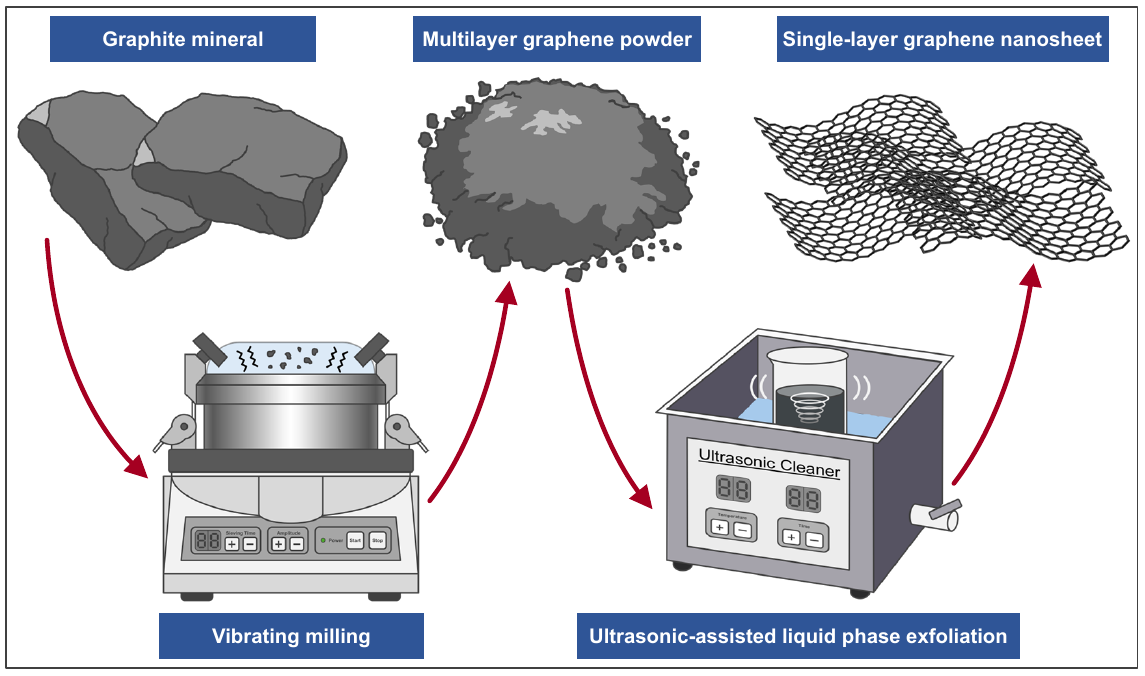} \\
\medskip
\noindent{\textbf{Caption of Graphical Abstract:} A schematic representation of the novel ultrasonic exfoliation method demonstrates the enhanced separation and handling of high-quality graphene, facilitating scalable production and applications.}
\bigskip
\label{GA}
\end{graphicalabstract}

\begin{highlights}
\item Developed a novel ultrasonic intercalation exfoliation method to synthesize high-crystallinity monolayer graphene.
\item Graphene nanosheets exhibited thickness 0.5 nm and micro-sized dimensions, confirmed by AFM and SEM.
\item XRD and Raman analysis revealed the layered and turbostratic structure of the prepared graphene.
\item XPS data showed uniform deposition of graphene on copper foil, altering the surface chemical bonding.
\item Approach enables scalable production of high-quality graphene, enhancing its applications across diverse fields.
\end{highlights}


\end{frontmatter}


\section{Introduction}

Graphene is a two-dimensional crystalline material composed of a single layer of carbon atoms bonded through $sp^2$ hybrid orbitals in a honeycomb lattice structure. This unique atomic arrangement endows graphene with exceptional physical properties. Its densely packed carbon atoms and hexagonal lattice make it extraordinarily strong—over 200 times stronger than steel per unit weight \cite{yu2000strength}. Despite this strength, graphene exhibits high flexibility, allowing it to bend or deform without compromising its structural integrity \cite{lee2009elastic}. Its flat band electronic structure enables electrons to traverse the lattice at exceedingly high speeds, conferring exceptional electrical conductivity \cite{novoselov2004electric}. Additionally, graphene demonstrates remarkable performance in thermal conductivity \cite{balandin2008superior}, transparency \cite{wang2008transparent}, lightness \cite{hu2013ultralight}, and impermeability \cite{bunch2008impermeable}. These exceptional properties have spurred extensive research into potential applications across diverse fields such as electronics, spintronics, twistronics \cite{wang2016synthesis, Li2023}, energy storage, quantum science, biomedical devices, and sensors, etc (as summarized in Table~S1). To fully leverage these properties, understanding and optimizing graphene synthesis methods is crucial.

\begin{figure*}[t]
\centering
\includegraphics[width = 0.68\textwidth] {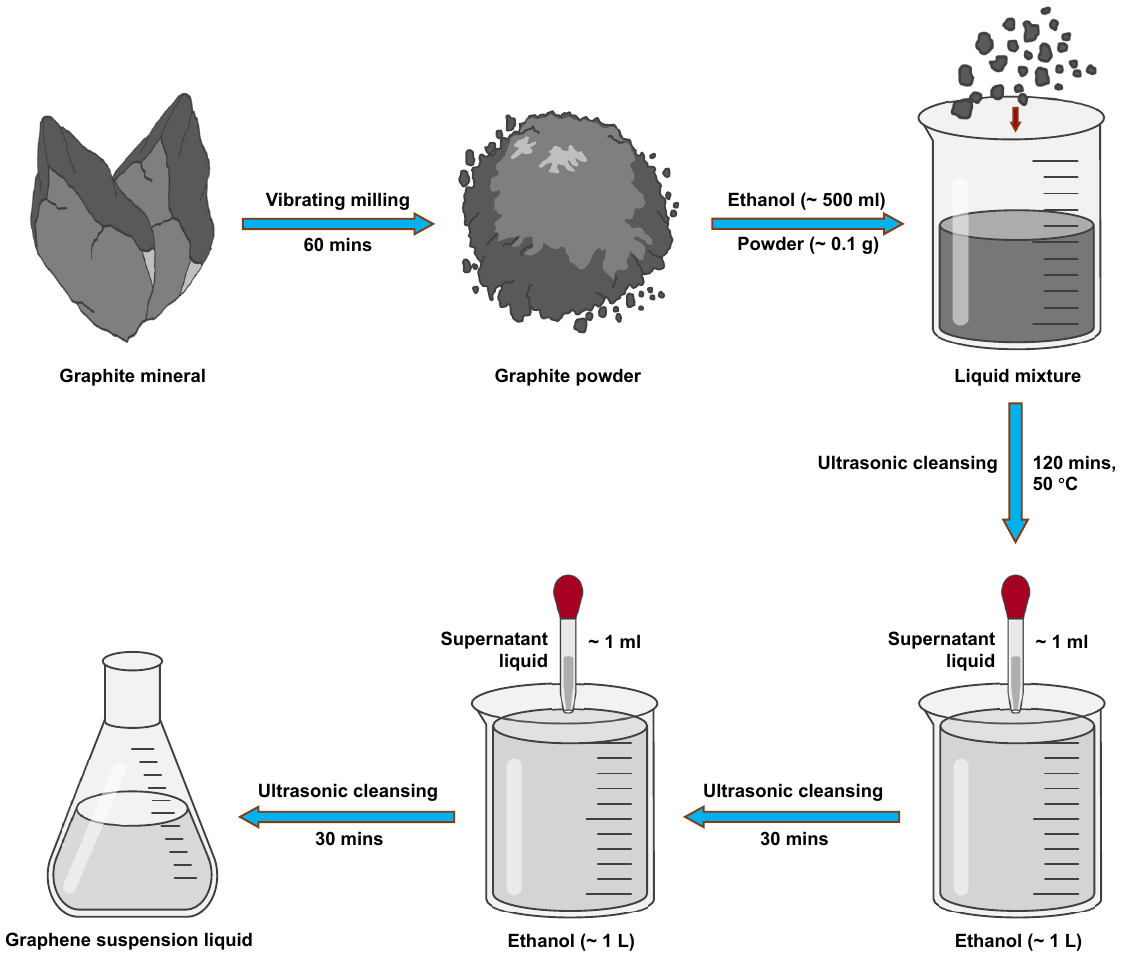}
\caption{A schematic representation of the synthesis procedure for a monolayer graphene suspension in liquid.}
\label{GraphicalAbstract}
\end{figure*}

The breakthrough achievement in isolating single-layer graphene originated from the pioneering work of Andre Geim and Konstantin Novoselov in 2004, utilizing a method involving the repeated peeling of bulk graphite using adhesive tape \cite{novoselov2004electric}. This milestone contribution to graphene science earned them the Nobel Prize in Physics in 2010. Subsequently, several methods for graphene synthesis emerged. The chemical vapor deposition (CVD) technique involves the decomposition of carbon-containing gases, such as methane, on a heated metal substrate, enabling the formation of high-quality graphene. This method exhibits scalability, proving advantageous for industrial-scale production \cite{kim2009large}. Similarly, the plasma-enhanced CVD (PECVD) method utilizes plasma to govern the growth rate and properties of graphene during the decomposition of carbon-containing gases on metal substrates \cite{zafar2022plasma}. The epitaxial growth method involves heating silicon carbide (SiC) crystals in a controlled environment, causing the sublimation of silicon atoms and leaving carbon atoms to epitaxially grow graphene on the SiC substrate \cite{strupinski2011graphene}. While epitaxial growth yields high-quality graphene, it faces limitations due to the availability and expense of high-quality SiC substrates. Chemical reduction of graphene oxide through high-temperature treatments or chemical agents results in reduced graphene oxide, eliminating oxygen-containing functional groups \cite{salas2010reduction}. This process may alter physical properties and introduce defects on the nanosheets. Liquid-phase exfoliation methods utilize ultrasonication to exfoliate bulk graphite into layered graphene flakes within solvents or surfactants \cite{lotya2009liquid}. Electrochemical exfoliation applies an electric potential to a graphite electrode in an electrolytic solution, leading to the exfoliation of graphite into graphene \cite{wang2011high}. These exfoliation techniques streamline graphene preparation, enabling easier isolation in solution compared to extraction from substrates. The detailed comparison among various synthesis methods is listed in Table~\ref{Syn-Meths}. Among these methods, liquid-phase exfoliation stands out due to its simplicity and effectiveness.

\begin{figure*}[t]
\centering
\includegraphics[width = 0.68\textwidth] {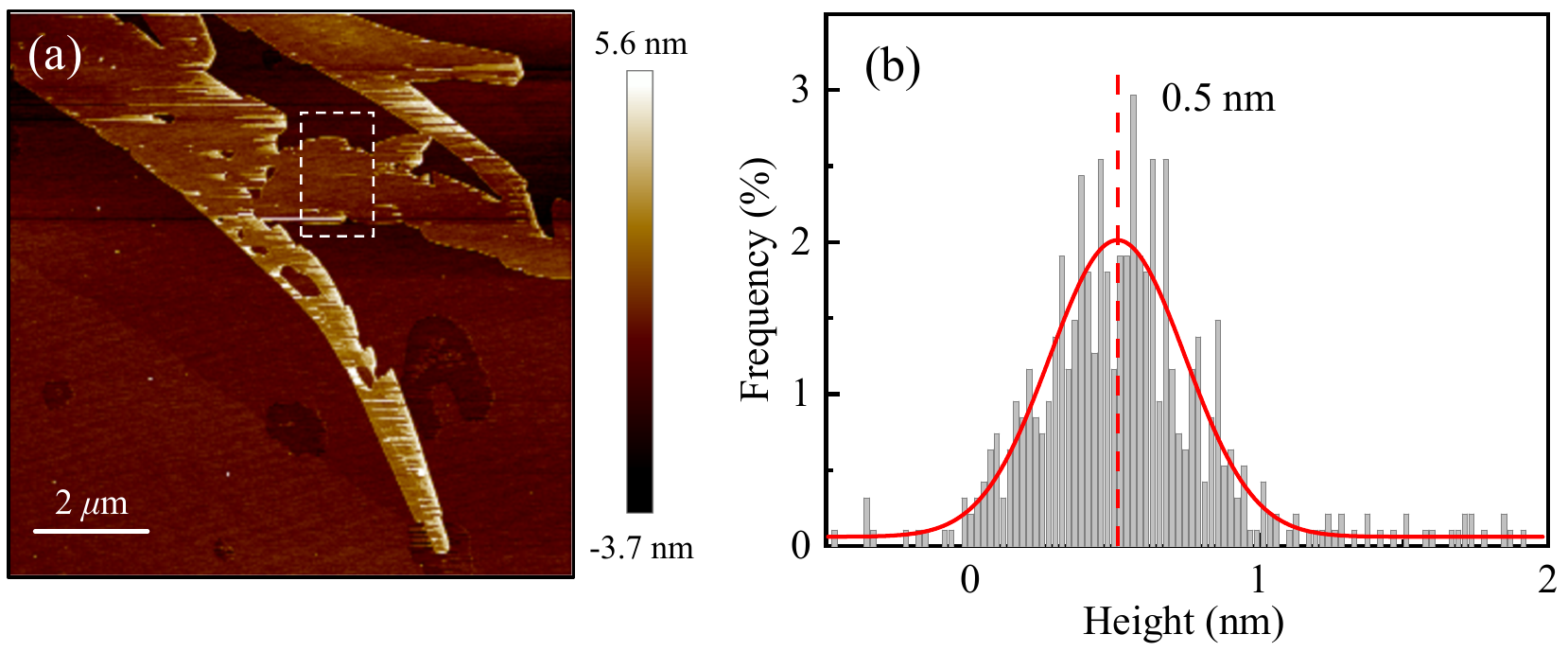}
\caption{(a) Topography of an AFM image in a 10 $\mu$m $\times$ 10 $\mu$m area, with a scale bar of 2 $\mu$m. (b) Measured height in the marked box in (a), revealing a measured height of $\sim$ 0.5 nm.}
\label{AFM}
\end{figure*}

Among the various methods for preparing graphene, liquid-phase exfoliation is a relatively simple and effective technique. Ultrasound-assisted exfoliation utilizes the cavitation effect, which creates high pressure along the interlayers of graphite, rapidly exfoliating it to form graphene with few or single layers \cite{gu2019method}. By applying low ultrasonic intensity over a long duration, graphene with few defects and a complete structure can be obtained after multiple cycles of ultrasonic treatment and centrifugation \cite{parvez2015exfoliation}. However, this method is time-consuming, and the graphene yield is very low, making it unsuitable for large-scale production. In this study, we applied ball milling to graphite raw materials during the precursor stage \cite{Li2007-1, Li2007-2, Wu2020} before exfoliation and incorporated extra heating during the ultrasonic exfoliation process. This approach effectively reduced the time required and increased the yield of graphene, thereby improving the efficiency of graphene production, making it more suitable for engineering applications.

In this paper, we employed the water-bath heating ultrasonic exfoliation method to disperse graphene in ethanol without altering its chemical properties. The resulting graphene suspension was extensively characterized using atomic force microscopy (AFM), scanning electron microscopy (SEM), X-ray diffraction (XRD), Raman spectroscopy, and X-ray photoelectron spectroscopy (XPS). AFM and SEM were used to study the topography and morphology of the monolayer graphene, while XRD and Raman spectroscopy were employed to analyze its structural phase and turbostratic stacking nature. Additionally, XPS was utilized to determine the bonding type of graphene and the surface valence change of the copper foil after graphene coating. The successful preparation of graphene through the ultrasonic exfoliation method offers a promising means to boost graphene yield. Furthermore, the evident alterations brought about by graphene coating hold significant potential for enhancing its industrial applications.

\begin{figure*}[t]
\centering
\includegraphics[width = 0.68\textwidth] {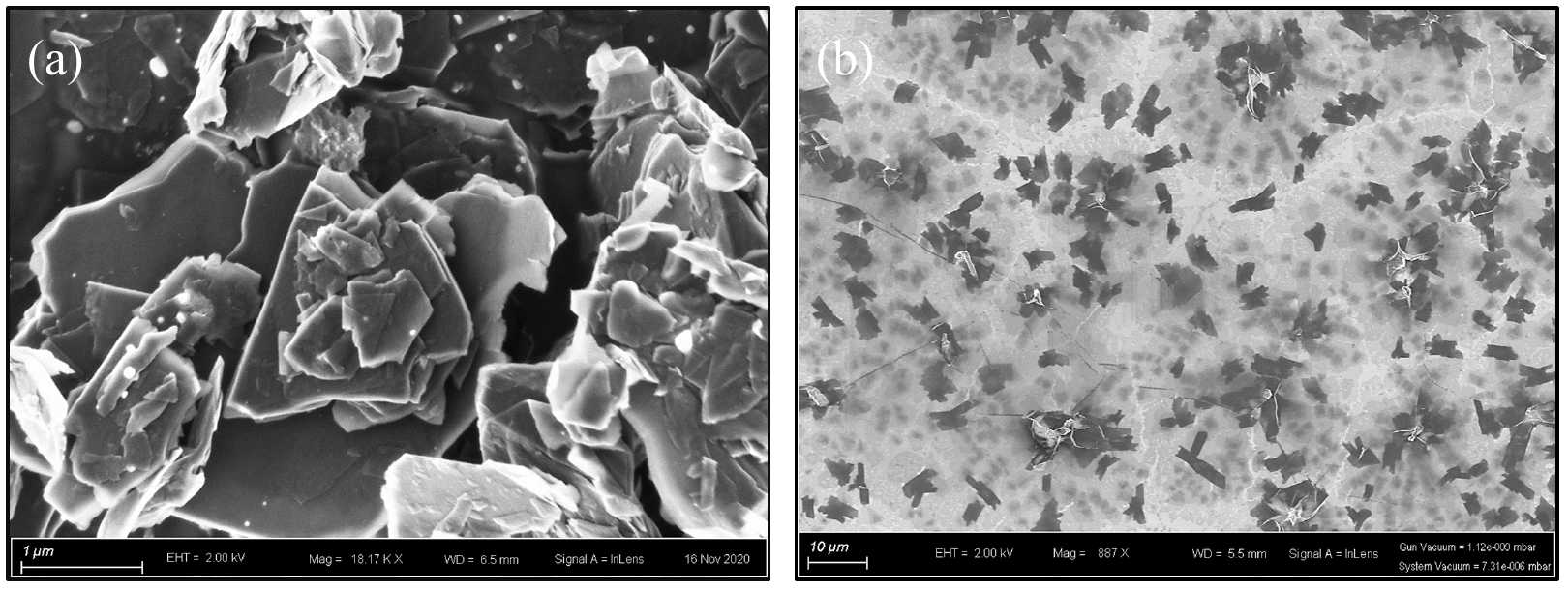}
\caption{(a) Microstructure of graphite powders, with a scale bar of 1 $\mu$m. (b) Top view of graphene suspension on aluminum foil, with a scale bar of 10 $\mu$m.}
\label{SEM}
\end{figure*}

\begin{figure}[t]
\centering
\includegraphics[width = 0.48\textwidth] {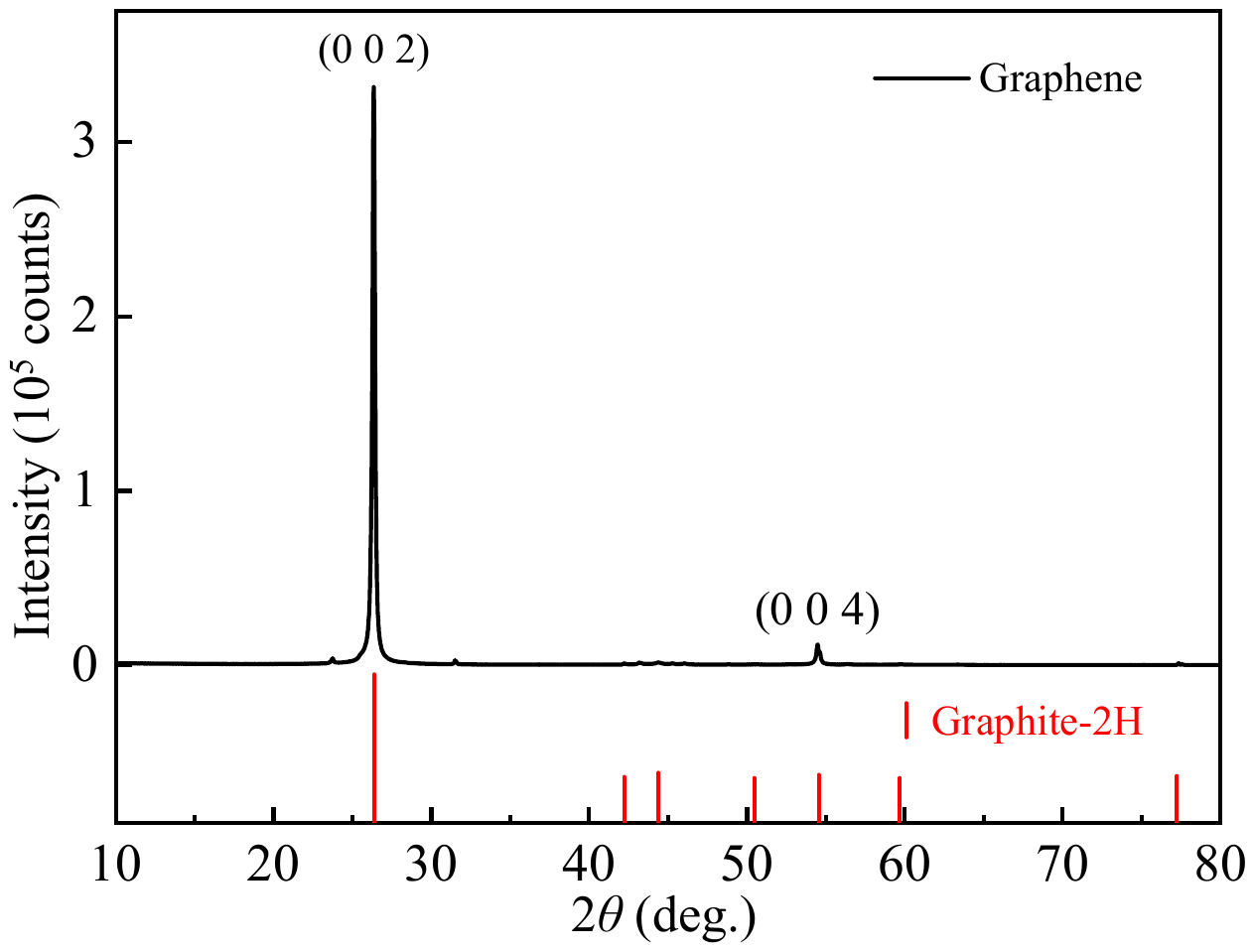}
\caption{The observation of a preferred orientation of (0 0 \textit{h}) in graphite-2H phase is evident in graphene.}
\label{XRD}
\end{figure}

\begin{figure*}[t]
\centering
\includegraphics[width = 0.68\textwidth] {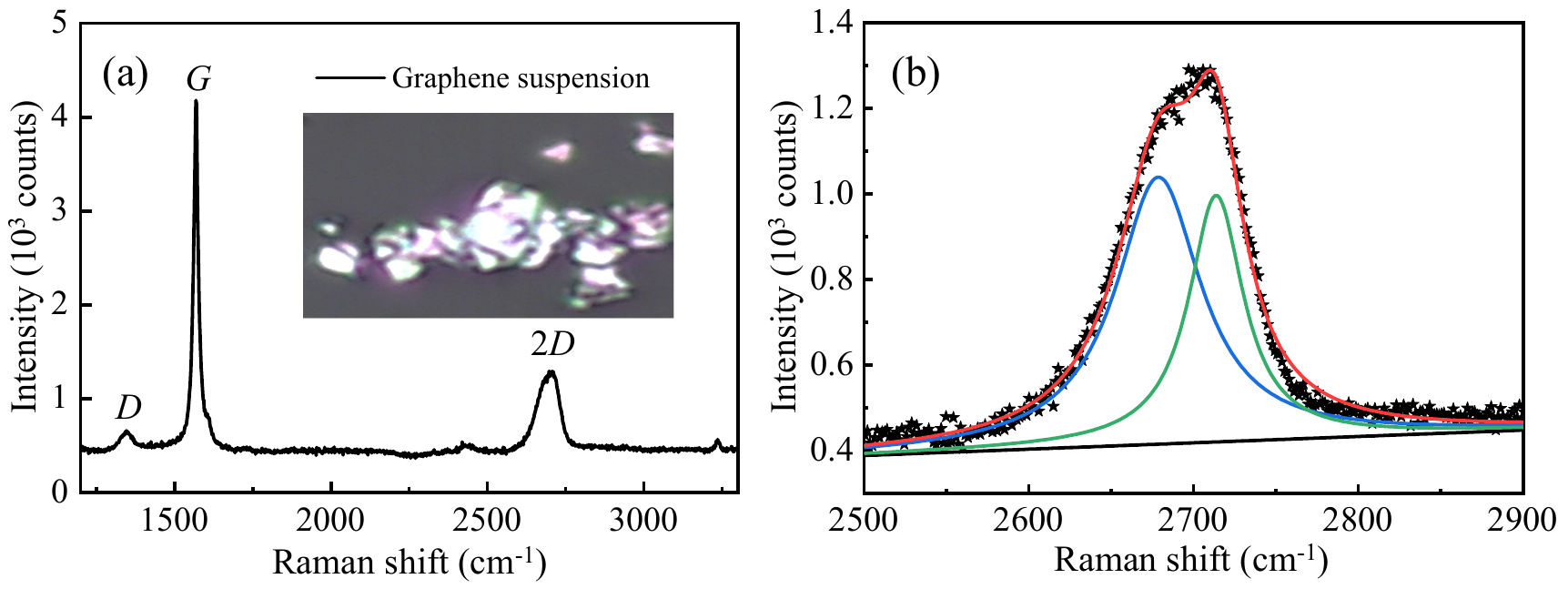}
\caption{(a) Under an excitation source of 514 nm. (b) Fitting of Lorentz functions for the 2\textit{D} band.}
\label{Raman}
\end{figure*}

\begin{figure*}[t]
\centering
\includegraphics[width = 0.68\textwidth] {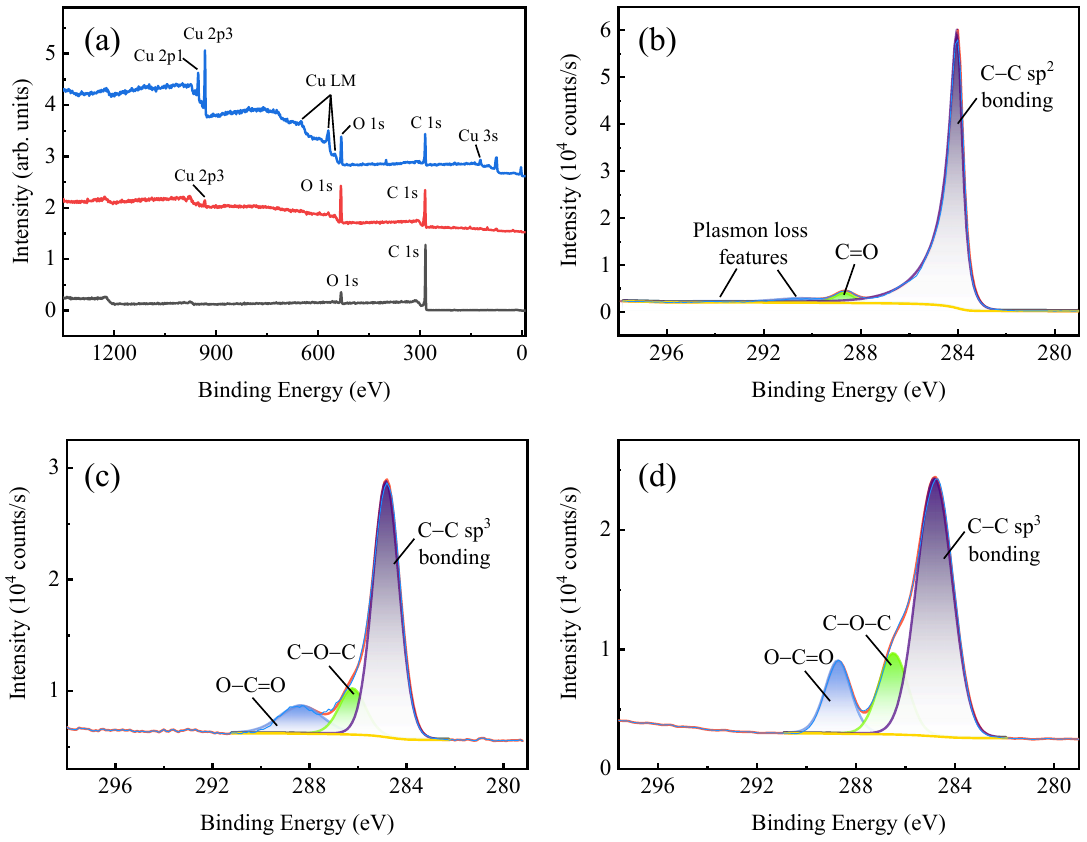}
\caption{(a) Survey XPS spectra depicting graphene powder (black line), graphene suspension on Cu foil substrate (red line), and copper foil (blue line). XPS C 1\emph{s} core level spectra of (b) graphene powder, (c) copper foil, and (d) graphene suspension on the Cu foil substrate.}
\label{XPS}
\end{figure*}

\section{Materials and methods}

\subsection{Raw materials}

There are various types of raw graphite materials, such as natural graphite mineral and chemically synthesized graphite like reduced graphene oxide. The physical and chemical properties of graphene differ significantly between these materials. For our study, we utilized high-quality natural graphite mineral as the primary raw material to fabricate monolayer graphene.

\subsection{Synthesis procedures}

To synthesize monolayer graphene, we utilized natural graphite mineral and ethanol (GR, Macklin). The procedure for preparing monolayer graphene is illustrated in Fig.~\ref{GraphicalAbstract}. The natural graphite mineral was subjected to a 60-minute milling process using a Vibratory Micro Mill (FRITSCH PULVERISETTE 0) \cite{Li2007-1, Li2007-2, Wu2020}, which effectively pulverized the bulk graphite into fine graphite powders. A total of 0.1 g of graphite powder was dispersed in 500 mL of GR ethanol. The resulting mixture was sealed in a glass bottle and placed in a water-bath ultrasonic cleaner (40 KHz, 300 W), where it was gradually heated to 50 °C over a period of 120 minutes.

After ultrasonic treatment, the bottle was left undisturbed for 60 minutes to allow heavy impurities to settle. Subsequently, 1 mL of the supernatant liquid was extracted and added to a beaker containing 1000 mL of GR ethanol. The mixture underwent an additional 30 minutes of ultrasonic treatment while being stirred simultaneously. These steps facilitated the penetration of ethanol and microbubbles into the graphite interlayers, transforming the bulk graphite into a homogeneous monolayer graphene suspension.

\subsection{Characterizations}

The thickness and number of layers in graphene were assessed using AFM images obtained with a Nanoscope MultiMode instrument (Digital Instruments/Bruker Systems). To prepare the AFM samples, the graphene suspension was deposited onto a mica substrate and dried under an incandescent lamp for 5 minutes before measurement. The microstructure of both raw graphite powders and the graphene suspension were examined using a SEM (Zeiss Sigma FESEM) operated at an accelerating voltage of 2.0 kV. For the SEM experiment, a droplet of graphene suspension was deposited onto a piece of aluminum foil and dried under an incandescent lamp before sample mounting. The diffraction pattern of graphene was captured using an X-ray diffractometer (Rigaku SmartLab 9 kW) operated at 45 kV and 200 mA. Large amounts of dispersed graphene powders were utilized as samples for these measurements. The Raman spectra were generated using a 514 nm excitation source and captured via a micro-Raman spectrometer (Horiba LABHRev-UV). The binding energies of C-C bonds and their electron configuration in graphene were determined through XPS (Thermo Fisher Scientific ESCALAB 250Xi). A full-scan spectrum was obtained at an incident energy of 50.0 eV, while a narrow-scan spectrum was recorded at 20.0 eV, with a step size of 0.1 eV.

\section{Results and discussion}

\subsection{Determining graphene thickness and layer number via atomic force microscopy}

AFM is commonly used to determine the thickness of 2D nanomaterials. Previous literature reports the thickness of monolayer graphene on a mica substrate ranging from 0.4 to 0.9 nm \cite{xu2010graphene}. The variation in graphene thickness is attributed to interactions between the AFM probe and graphene, influenced by substrate surface energy, graphene structure, and sample preparation \cite{shearer2016accurate}. In our study, thickness measurements were conducted using AFM tapping mode. Fig.~\ref{AFM}(a) illustrates multiple layers of graphene nanosheets, with the optimal area measuring $\sim$ 0.5 nm in height as shown in Fig.~\ref{AFM}(b). Additionally, AFM images indicate that the graphene nanosheets exhibit micro-sized dimensions in length.

\subsection{Microstructural analysis using scanning electron microscopy}

The microstructural disparities between graphite and graphene were investigated using SEM \cite{Wu2020}. In Fig.~\ref{SEM}(a), clusters of micro-sized graphene powders are depicted, exhibiting thicknesses of several hundred nanometers. These particles exhibit cross-sectional views showcasing uniformly-stacked layered structures connected through van der Waals interactions. However, the particle size distribution appears non-uniform, with smaller flakes interspersed among larger particles. Contrastingly, Fig.~\ref{SEM}(b) illustrates graphene nanosheets distributed separately on aluminum foil, showcasing an average particle size of $\sim$ 5 $\mu$m. Some graphene nanosheets are observed to stack together with dislocations evident between each layer, indicating a lack of interaction between the layers. Notably, the ultrasonic treatment effectively exfoliates graphite particles into graphene nanosheets, overcoming inter-layer forces in graphite.

\subsection{Phase identification through x-ray diffraction}

The structural phase of ultrasonically exfoliated graphene was investigated using XRD. As shown in Fig.~\ref{XRD}, distinct Bragg peaks of (0 0 2) and (0 0 4) were identified in the graphene XRD pattern. The strong preferred orientation perpendicular to the (0 0 2) crystal face underscores the layered nature characteristic of two-dimensional materials. The peak position for (0 0 2) is 2$\theta$ = 26.34$^{\circ}$, with an incident X-ray beam wavelength of $\lambda$($\textit{K}_{\alpha1}$) = 1.540593 {\AA}. These findings strongly indicate the efficiency of the ultrasonic exfoliation method in obtaining high-crystallinity graphene.

\subsection{Raman spectroscopy}

The Raman spectrum of the graphene suspension was obtained under consistent conditions: a 514 nm excitation source, 4 s per cycle, and 2 cycles. Fig.~\ref{Raman}(a) exhibits three prominent peaks: the \textit{D} peak located at 1345.8 cm$^{-1}$, the \textit{G} peak at 1568.8 cm$^{-1}$, and an approximate 2\textit{D} band around 2700 cm$^{-1}$. The \textit{D} peak arises from the second-order zone-boundary phonons in defected graphite \cite{nemanich1979first}, indicative of observations at the boundary of graphene layers. The \textit{G} peak, originating from the double degenerate zone center \textit{E}$_{2g}$ mode \cite{tuinstra1970raman}, is 11.2 cm$^{-1}$ higher than the bulk graphite at 1580 cm$^{-1}$, attributed to chemical doping. The 2\textit{D} band, associated with the second-order zone-boundary phonons \cite{ferrari2006raman, vidano1981observation}, is the second most prominent peak typically observed in graphite samples.

In Fig.~\ref{Raman}(b), the 2\textit{D} band is fitted using two Lorentz functions at 2678.7 cm$^{-1}$ and 2713.7 cm$^{-1}$, with full width at half maximum (FWHM) values of 62.7 cm$^{-1}$ and 40.2 cm$^{-1}$, respectively. The 2\textit{D} band's shape resembles neither few-layer graphene nor graphite samples \cite{ferrari2006raman, malard2009raman}; instead, it appears more akin to a turbostratic graphene sample, evident from its much smaller \textit{I}$_{2D}$/\textit{I}$_{G}$ and broader linewidth of the 2\textit{D} band compared to monolayer graphene \cite{malard2009raman, kokmat2023growth}. The optical microscopic image in the insert of Fig.~\ref{Raman}(a) clearly reveals the random stacking of graphene layers along the \textit{c} axis.

\subsection{X-ray photoelectron spectroscopy}

The bonding characteristics of the graphene sample were investigated using XPS spectra \cite{ZHOU2023101248}. Fig.~\ref{XPS}(a) displays elemental analyses of graphene powder (black line), graphene suspension on Cu foil substrate (red line), and copper foil (blue line). The graphene powder sample shows peaks solely for C and O elements across the entire binding energy range, with the C 1\emph{s} peak notably stronger than the O 1\emph{s} peak. However, in the deposited graphene suspension sample, the intensity of the O 1\emph{s} peak significantly increases, accompanied by a small Cu 2$p^3$ peak. The heightened O 1\emph{s} peak predominantly originates from O atoms in the copper foil substrate, while the appearance of Cu 2$p^3$ primarily stems from the Cu substrate. Considering XPS spectra collect elemental information only from the surface of samples, the prominence of the Cu 2$p^3$ peak compared to the Cu sample is subtle. Moreover, other peaks linked to the Cu element, including Cu 2$p^1$, Cu 3\emph{s}, and Cu LMM, are absent in the deposited graphene suspension. This indicates the uniform deposition of a thin layer of layered graphene on copper foil surface, as observed in Fig.~\ref{SEM}(b).

\begin{table*}[th]
\small
\caption{Comparison of diverse fabrication methods for graphene. ME = Mechanical exfoliation. CVD = Chemical vapor deposition. LPE = Liquid-phase exfoliation. COR = Chemical oxidation-reduction. CRGO = Chemical reduction of graphene oxide. EG = Epitaxial growth. EE = Electrochemical exfoliation. AD = Arc discharge. PECVD = Plasma-enhanced CVD. FRGO = Flash reduction of graphene oxide. TS = this study.}
\label{Syn-Meths}
\setlength{\tabcolsep}{6.6mm}{}
\renewcommand{\arraystretch}{1.1}
\begin{tabular}{l|l|ll}
\hline
\hline
Method                                        &Advantage                                                    & Disadvantage                                                & Refs.                                     \\ [2pt]
\hline
ME                                            &$\bullet$ High quality $\bullet$ Direct access               &$\bullet$ Small scale $\bullet$ Low production               &\cite{yi2015review}                        \\
CVD                                           &$\bullet$ Scalable $\bullet$ Direct growth                   &$\bullet$ High temperature                                   &\cite{zhang2013review}                     \\
                                              &$\bullet$ Controlled thickness \& morphology                 &$\bullet$ Possible impurities                                &                                           \\
LPE                                           &$\bullet$ Simple \& scalable                                 &$\bullet$ Few layer                                          &\cite{hernandez2008high}                   \\
                                              &$\bullet$ Production of dispersions                          &$\bullet$ Requiring stable agents                            &                                           \\
COR                                           &$\bullet$ Widely used $\bullet$ Facile integration           &$\bullet$ Harsh oxidation conditions                         &\cite{zhao2014interactive}                 \\
                                              &$\bullet$ Improved compatibility                             &$\bullet$ Post treatments                                    &                                           \\
CRGO                                          &$\bullet$ Simple \& low cost $\bullet$ Scalable              &$\bullet$ Reduced conductivity $\bullet$ Defects             &\cite{chua2014chemical}                    \\

EG on SiC                                     &$\bullet$ Direct growth $\bullet$ High quality               &$\bullet$ Limited to specific SiC                            &\cite{mishra2016graphene}                  \\
                                              &$\bullet$ Controlled thickness                               &                                                             &                                           \\
EE                                            &$\bullet$ Simple $\bullet$ Tunable properties                &$\bullet$ Small scale                                        &\cite{liu2019synthesis}                    \\
                                              &$\bullet$ Environmentally friendly                           &$\bullet$ Reduced crystallinity                              &                                           \\
AD                                            &$\bullet$ High quality $\bullet$ Large sheets                &$\bullet$ Low yield \& scalability                           &\cite{subrahmanyam2009simple}              \\
PECVD                                         &$\bullet$ Scalable $\bullet$ Large sheets                    &$\bullet$ Impurities \& defects                              &\cite{li2016controllable}                  \\
                                              &$\bullet$ Controlled thickness \& uniformity                 &$\bullet$ Specific growth conditions                         &                                           \\
FRGO                                          &$\bullet$ Rapid reduction                                    &$\bullet$ Limited-controlled properties                      &\cite{park2015environmentally}             \\
                                              &$\bullet$ Minimal use of reducing agents                     &$\bullet$ Reduced conductivity                               &                                           \\
Coating                                       &$\bullet$ Simple $\bullet$ Uniformity                        &$\bullet$ Limited precise control                            &TS                                         \\
                                              &$\bullet$ Scalability $\bullet$ Versatility                  &$\bullet$ Substrate sensitivity $\bullet$ Defects            &                                           \\
                                              &$\bullet$ Controlled thickness                               &$\bullet$ Wasteful use of materials                          &                                           \\
                                              &                                                             &$\bullet$ Post treatments required                           &                                           \\
                                              \hline
\hline
\end{tabular}
\end{table*}

Figs.~\ref{XPS}(b)-\ref{XPS}(d) exhibit the C 1\emph{s} spectra of graphene powder, copper foil, and graphene suspension on the copper foil substrate. As shown in Fig.~\ref{XPS}(b), an asymmetric $sp^2$ peak is observed at 284.0 eV, accompanied by small satellite peaks at 290.38 eV and 293.75 eV, indicating a high concentration of the $sp^2$ bonding. Additionally, a C=O peak is visible at 288.65 eV, originating from minimal oxidation impurities. These spectral features are typical of samples with a high concentration of graphene-like structures, confirming the high purity of the graphene powder.

Conversely, as shown in Fig.~\ref{XPS}(c), the copper foil sample displays an $sp^3$ bonding fitted with a symmetric Gaussian-Lorentzian peak at 284.8 eV, 0.8 eV higher than that of the $sp^2$ bonding. Notably, contributions from carbon-oxygen bonds (including C--O--C at 286.19 eV and O--C=O at 288.32 eV) are much larger than in the graphene powder sample. This $sp^3$ bonding pattern, featuring C--O--C and O--C=O bonds, typifies a detectable quantity of adventitious carbon contamination on the copper foil surface due to atmospheric exposure before measurements.

Similarly, as shown in Fig.~\ref{XPS}(d), the deposited graphene suspension sample exhibits analogous adventitious carbon contamination, submerging the asymmetric $sp^2$ peak within the broad $sp^3$ peak. However, the contribution of the C=O peak from the deposited graphene remains discernible through the increased intensity of the O--C=O peak. These findings clearly demonstrate that even a small amount of graphene layers deposited on metal significantly alters the chemical bonds on the metal surface, thereby modifying the metal sample's electrical transport properties. This indicates that depositing graphene onto material surfaces, instead of doping graphene into materials, can alter the chemical valence states of materials. The XPS spectra characterization results present a novel method by which graphene can be utilized to adjust the properties of materials.

\section{Conclusion}

We produced monolayer graphene nanosheets using an advanced heat-assisted high-energy ultrasonic exfoliation method. Various laboratory methods were employed to characterize the properties of the extracted graphene sample. AFM characterizations revealed that our prepared monolayer graphene sample has a measured thickness of $\sim $ 0.5 nm, considering the wrinkle in freestanding monolayer graphene. SEM characterizations displayed the morphology of both graphite and graphene, clearly illustrating the layered structure of graphene nanosheets. XRD patterns exhibited a strong (0 0 2) peak in the graphene powder samples, highlighting the 2D nature of the graphene sample. The Raman spectrum provided evidence that the detectable components in our graphene suspension sample were attributed to turbostratic graphene with distinct interlayer spacing. Surveying XPS spectrum and C 1\emph{s} core level spectra unveiled the deposition of graphene layers on copper metal foil, revealing variations in chemical bonding.

By enhancing the ultrasonic power and elevating the water bath temperature during the preparation process, we identified an efficient method to synthesize a significant amount of monolayer graphene. Monolayer graphene can significantly alter the surface chemical valence state of metal, offering insights into manipulating the transport properties of conductive materials. Future studies could focus on investigating graphene-harmonic interactions on electrodes.

\section*{CRediT authorship contribution statement}

\textbf{Kaitong Sun:} Writing - original draft, Visualization, Methodology, Investigation, Formal analysis, Data curation, Conceptualization.
\textbf{Si Wu:} Visualization, Methodology, Investigation, Formal analysis, Data curation, Conceptualization.
\textbf{Junchao Xia:} Visualization, Methodology, Investigation, Formal analysis, Data curation, Conceptualization.
\textbf{Yinghao Zhu:} Visualization, Methodology, Investigation, Formal analysis, Data curation, Conceptualization.
\textbf{Guanping Xu:} Visualization, Methodology, Investigation, Formal analysis, Data curation, Conceptualization.
\textbf{Hai-Feng Li:} Writing - review \& editing, Visualization, Supervision, Project administration, Methodology, Funding acquisition, Conceptualization.

\section*{Declaration of competing interest}

The authors declare that they have no known competing financial interests or personal relationships that could have appeared to influence the work reported in this paper.

\section*{Data availability}

Data will be made available on request.

\section*{Acknowledgements}

This work was supported by the Science and Technology Development Fund, Macao SAR (File Nos. 0090{/}2021{/}A2 and 0104{/}2024{/}AFJ) and University of Macau (MYRG{-}GRG2024{-}00158{-}IAPME).

\bibliographystyle{elsarticle-num-names}
\bibliography{GR}

\clearpage

\begin{table*}[ht]
\small
\caption{Summary of various applications of graphene.}
\label{app}
\setlength{\tabcolsep}{8.1mm}{}
\renewcommand{\arraystretch}{1.1}
\begin{tabular}{l|ll}
\hline
\hline
Applicaiton                 & Remark                                                                                               & Refs.                                                                   \\ [2pt]
\hline
Electronics                 &$\bullet$ High conductivity for electronic sensors and devices                                        &\cite{geim2007rise, novoselov2004electric}                               \\
Energy storage              &$\bullet$ Supercapacitor $\bullet$ Anode materials for batteries                                      &\cite{zhang2009carbon, al2019emergence}                                  \\
Composite materials         &$\bullet$ Improving mechanical, electrical \& thermal properties                                      &\cite{lim2014improved, kuilla2010recent}                                 \\
Sensors                     &$\bullet$ Gas \& chemical $\bullet$ Biosensors \& wearable devices                                    &\cite{schedin2007detection, kim2017graphene}                             \\
Biomedical                  &$\bullet$ Drug delivery $\bullet$ Imaging and diagnostic tools                                        &\cite{yang2016stimuli, garg2015graphene}                                 \\
                            &$\bullet$ Biocompatible neural probes $\bullet$ Brain-machine interfaces                              &                                                                         \\
Automotive industry         &$\bullet$ Lightweight \& durable materials                                                            &\cite{elmarakbi2018state}                                                \\
Thermal management          &$\bullet$ Electric vehicles $\bullet$ Interface                                                       &\cite{liu2020graphene, shahil2012thermal}                                \\
Aerospace                   &$\bullet$ Structural materials $\bullet$ Lightning strike protection                                  &\cite{monetta2015graphene, raimondo2018multifunctional}                  \\
Environmental               &$\bullet$ Water purification $\bullet$ Air filtration $\bullet$ Monitoring pollutants                 &\cite{dervin20162d, zhang2019graphene}                                   \\
                            &$\bullet$ Capturing and immobilizing pollutants from soil                                             &                                                                         \\
                            &$\bullet$ Sound{/}oil absorption $\bullet$ Reduce impact of electronic waste                          &                                                                         \\
Coatings \& corrosion       &$\bullet$ Corrosion protection $\bullet$ Anti-fouling coating                                         &\cite{kirkland2012exploring, seo2018anti}                                \\
Construction                &$\bullet$ Concrete reinforcement $\bullet$ Smart material                                             &\cite{chuah2014nano, yu2017graphene}                                     \\
Photodetector               &$\bullet$ High speed $\bullet$ Broadband                                                              &\cite{mueller2010graphene, liu2014graphene}                              \\
Quantum science             &$\bullet$ Quantum dots \& qubits $\bullet$ Quantum dot integration                                    &\cite{wang2010graphene, li2019review}                                    \\
                            &$\bullet$ Quantum sensing $\bullet$ Quantum Hall effect                                               &                                                                         \\
Lubricants \& friction      &$\bullet$ Solid lubricants  $\bullet$ Tribological applications                                       &\cite{berman2014graphene, chouhan2020surface}                            \\
Agriculture                 &$\bullet$ Nanoparticle delivery $\bullet$ Plant growth enhancement                                    &\cite{liu2013graphene, chakravarty2015graphene}                          \\
3D Printing                 &$\bullet$ Enhanced filaments $\bullet$ Structural components                                          &\cite{kotsilkova2019exploring, you2021structural}                        \\
Cosmetics \& skincare       &$\bullet$ Sunscreen formulations $\bullet$ Anti-aging products                                        &\cite{meftahi2022nanocelluloses, bhat2022recent}                         \\
Superconductivity           &$\bullet$ Enhancing critical current density                                                          &\cite{gonzalez2008critical, feigel2008proximity}                         \\
                            &$\bullet$ Proximity-induced superconductivity                                                         &                                                                         \\
Catalysis                   &$\bullet$ Electrocatalysts $\bullet$ Photocatalysis                                                   &\cite{xia2014recent, li2016graphene}                                     \\
\hline
\hline
\end{tabular}
\end{table*}

\end{document}